\documentclass[twocolumn,showpacs,aps,prb]{revtex4}
\usepackage{graphicx}

\begin{document}
\textheight 10.0in

%\draft                           % Comment this command for twocolumn layout

%\wideabs                       % Uncomment this command for twocolumn layout
                                 % an also the line indicated after PACS.

\title{Analysis of electric-field-induced spin splitting in wide modulation-doped quantum wells}

\author{U. Ekenberg}
\altaffiliation{Corresponding author. ulfe@kth.se}
\affiliation{School of Information and Communication Technology,
Royal Institute of Technology, SE-16440 Kista, Sweden}
\author{D.M. Gvozdi\'{c}}
\affiliation{Faculty of Electrical Engineering, University of
Belgrade, P.O. Box 35-54, 11120 Belgrade, Serbia}
\date{\today}

\begin{abstract}

We analyze the proper inclusion of electric-field-induced spin
splittings in the framework of the envelope function approximation.
We argue that the Rashba effect should be included in the form of a
macroscopic potential as diagonal terms in a multiband approach
rather than the commonly used Rashba term dependent on \emph{k} and
electric field. It is pointed out that the expectation value of the
electric field in a subband is sometimes not unique because the
expectation values can even have opposite signs for the spin-split
subband components. Symmetric quantum wells with Dresselhaus terms
and the influence of the interfaces on the spin splitting are also
discussed. We apply a well established multiband approach to wide
modulation-doped InGaSb quantum wells with strong built-in electric
fields in the interface regions. We demonstrate an efficient
mechanism for switching on and off the Rashba splitting with an
electric field being an order of magnitude smaller than the local
built-in field that determines the Rashba splitting. The
implications of our findings for spintronic devices, in particular
the Datta-Das spin transistor and proposed modifications of it, are
discussed.

\end{abstract}


\maketitle

%\newpage

\section{Introduction}

The interaction of a spin with a magnetic field or magnetic ions has
been studied extensively over the years. A surprisingly efficient
mechanism to produce a spin splitting is utilizing the combination
of inversion asymmetry and spin-orbit interaction.
\cite{Winklerbook} One convenient way to control the spin splitting
is the Rashba effect which results from structure inversion
asymmetry (SIA). \cite{Bychkov} An applied or built-in macroscopic
electric field is seen in the frame of a moving electron as having a
magnetic-field component and yields a spin splitting. In this way
one can utilize many of the mechanisms from conventional electronics
which is mainly controlled by electric fields. The idea of using the
spin of the carriers in addition to its charge has resulted in a
research area called spintronics. \cite{Zutic} Another spin
splitting mechanism resulting from the lack of inversion symmetry of
the zinc-blende lattice [bulk inversion asymmetry (BIA)] is called
the Dresselhaus effect. \cite{Dresselhaus} Both the Rashba and the
Dresselhaus effects are frequently included via terms linear in the
wave vector \emph{k}. However, they are the lowest-order terms of
more accurate expressions that are obtained from multiband envelope
function theory. \cite{Winklerbook}

In Sec. II we will recapitulate the foundations of the commonly used
envelope function approximation \cite{LuttingerKohn,Luttinger} in
order to set the ground for an analysis of the proper inclusion of
SIA and BIA within the framework of this approximation. We make here
the important distinction between slowly and rapidly varying
potentials. In Sec. III we apply the multiband theory to an
interesting system, wide \emph{n}-type modulation-doped (MD) quantum
wells (QWs) with strongly nonuniform electric fields. Here there are
strong built-in electric fields of opposite signs at the two
interfaces and one can expect a strong Rashba effect. It has
frequently been assumed that one needs a strong applied bias to get
a substantial Rashba splitting but interesting things can happen
also for small or moderate bias. We show how the built-in electric
fields in modulation-doped quantum wells can be utilized while
applying a much smaller external field. For very small asymmetry
interesting anticrossing phenomena occur. Furthermore, the spin
splitting due to the Dresselhaus effect in symmetric quantum wells
is found to be qualitatively different in modulation-doped quantum
wells compared to square wells. In Sec. IV we discuss the
implications of our results for spintronic devices, in particular
the Datta-Das spin field effect transistor. \cite{Datta} It has for
a long time been considered as a prototype of a spintronic device
but unfortunately the efforts to implement it in practice have not
yet been very successful. Finally, in Sec. V we discuss the results
and conclude.

\section{Theory}

The envelope function approximation (EFA) has been widely used
during several decades. Under the name effective-mass theory, it was
first applied to shallow-impurity states in bulk semiconductors.
\cite{Kittel,Kohn} The starting point is that the problem with the
band structure in the pure bulk material is assumed to have been
solved. According to Bloch's theorem the total wave function for
band $n$ is given by
\begin{equation}
\psi_{n\mathbf{k}} (\mathbf{r}) = e^{i \mathbf{k} \cdot \mathbf{r}}
u_{n\mathbf{k}}(\mathbf{r})
\end{equation}
where $u_{n\mathbf{k}}(\mathbf{r})$ has the periodicity of the
lattice. We then introduce a perturbation $U(\mathbf{r})$. An
essential assumption in the derivation is that it should be slowly
varying on the scale of the lattice constant. This assumption does
not always hold in the cases where the EFA has been applied. The
advantage of the EFA is that one can avoid the explicit inclusion of
the cell periodic potential. Only the slowly varying perturbation
$U(\mathbf{r})$ enters a Schr\"{o}dinger-like equation. With the
perturbation the total wave function can be expanded
\begin{equation}
\psi(\mathbf{r}) = \sum_n f_n(\mathbf{r})
u_{n\mathbf{k}}(\mathbf{r})
\end{equation}
where the summation is over all the energy bands. $f_n(\mathbf{r})$
is called an envelope function and the EFA gives a simple
prescription for the effective Hamiltonian operating on the envelope
function $f_n$ only, \cite{LuttingerKohn,Luttinger}
\begin{equation}
H = E(-i \nabla) + U(\mathbf{r})
\end{equation}

The kinetic-energy operator $E$ is obtained from the bulk band
structure $E(\mathbf{k})$. In general it is a matrix whose
eigenvalues gives the energy-band dispersion in the bulk. For the
perturbed problem $\mathbf{k}$ is replaced by $- i \nabla$, where
$\nabla$ is the gradient operator. In the case of a quantum well
grown along the $z$ direction, it is sufficient to replace $k_z$ by
the operator $- i \,
\partial/\partial z$ while $k_x$ and $k_y$ remain good quantum
numbers. The perturbation potential $U(\mathbf{r})$ is added along
the diagonal of the matrix.

For the impurity case the potential is not slowly varying in the
unit cell containing the impurity. For \emph{s} states having a
finite amplitude at the origin this sometimes gives important
deviations (often called central-cell corrections) from the
predictions of the EFA in its simplest form. \cite{Faulkner} The
ground state for a donor in Si is, for example, split up into three
levels where the energy separation between these levels can be
comparable to the predicted ground state energy. For \emph{p} and
\emph{d} states in Si the EFA works excellently and it also works
well for the ground state in direct-gap semiconductors.
\cite{Pantelides}

The EFA has also been used frequently for quantum well
heterostructures. Here the potential changes rapidly near the
interface between two materials. The range of this potential change
is of the order of the lattice constant. Thus, at first sight, it
appears that the EFA would not be applicable. In spite of this, the
agreement between its predictions and experimental results has
turned out to be quite good if the EFA is applied properly without
unnecessary approximations. The reason for this was examined by Burt
\cite{Burt} in a series of papers and led to a new set of boundary
conditions \cite{Foreman93,Foreman97} nowadays called the
Burt-Foreman boundary conditions. These boundary conditions were
first derived for the $6 \times 6$ Hamiltonian describing the
valence bands \cite{Foreman93} and later a prescription for
extending it to the $8 \times 8$ Hamiltonian including the
conduction band was given.\cite{Foreman97} Previously it was common
to symmetrize the order between operators and spatially varying
parameters but the Burt-Foreman boundary conditions were derived
with consideration of the cell-periodic wave functions
$u_{n\mathbf{k}}(\mathbf{r})$. Meney \emph{et al.} \cite{Meney}
analyzed different envelope function approaches and found support
for the Burt-Foreman boundary conditions. Burt's analysis also
explained why the EFA works quite well even for narrow quantum
wells. However, it should be kept in mind that caution is necessary
when applying the EFA to interface regions. In the EFA the
interfaces have usually been taken as abrupt steps. A rapidly but
continuously varying potential has also been considered by Stern and
Das Sarma \cite{Stern} but the influence on the subband energies was
found to be quite small.

The summation in Eq. (2) should in principle be over all the bands.
In practice one selects a finite number of important bands whose
interaction is included exactly in the matrix while the "remote"
bands are included perturbatively. \cite{Lowdin} A larger number of
bands included in the matrix gives an accurate description in a
larger \emph{k} range. A common choice that we apply in this paper
is to include the conduction, heavy-hole, light-hole, and split-off
bands in an $8 \times 8$ matrix. For a symmetric structure this
includes a twofold spin degeneracy.

For even more accurate results, $14 \times 14$ and $16 \times 16$
matrices have been considered.\cite{Fasol} Wissinger \emph{et al}.
\cite{Wissinger} have performed calculations for an asymmetric GaAs
quantum well both with the $8 \times 8$ and the $14 \times 14$
matrices, and compared to Raman-scattering experiments.
\cite{Richards} The deviation between the two models was rather
small and comparable to the deviation from the experimental results.
In the present case we consider InGaSb with a small band gap where
the $8 \times 8$ matrix is expected to be a good approximation. On
the other hand, it can be sufficient to use $6 \times 6$ matrices in
which the conduction band or the split-off band is among the remote
bands. The frequently applied Luttinger-Kohn Hamiltonian
\cite{LuttingerKohn,Luttinger} includes the heavy-hole and
light-hole bands in a $4 \times 4$ matrix. For a description of
electron subbands, it is convenient to use a one-band
(two-component) approximation in which all the other bands are
considered as remote and included via a modification of the
free-electron mass to an effective mass. As we will see below,
inclusion of spin effects in a one-band model leads to some
complications.

The recapitulation above of the essence of the EFA serves as the
basis for analyzing the inclusion of asymmetry-induced spin
phenomena. The inclusion of the Dresselhaus effect
\cite{Dresselhaus} seems clear. It stems from the microscopic
structure of the bulk material, influences the cell-periodic part
$u_{n\mathbf{k}}$ of the wave function, and results in a modified
bulk band structure. Thus it is appropriate to include it as
\emph{k}-dependent terms in the matrix which, after the replacement
$\mathbf{k} \to - i \nabla$, becomes the kinetic-energy operator.
Several terms of different order in $k$ enter the $8 \times 8$
Hamiltonian. \cite{Winklerbook} For electron subbands the
lowest-order term is linear in \emph{k}:
\begin{equation}
H_D = \beta (k_x \sigma_x - k_y \sigma_y),
\end{equation}
where $\sigma_x$ and $\sigma_y$ are Pauli matrices, and $\beta$ is a
material constant giving the strength of the Dresselhaus effect.

The inclusion of the Rashba effect \cite{Bychkov} is less
uncontroversial. It stems from a slowly varying macroscopic electric
field and, according to the principles of the EFA, it should be
included as a $z$-dependent potential along the diagonal of a matrix
of sufficient size. The Rashba effect in \emph{p}-channel Si
metal-oxide-semiconductor field-effect transistor (MOSFET)
structures was already included in this way in the 1970s and good
agreement with experiment was found. \cite{Bangert} Using a
multiband matrix for the kinetic energy, the inclusion of an
asymmetric potential results in a spin splitting for finite values
of the in-plane wave vector without inclusion of any special
\emph{k}-dependent terms [cf. Eq. (5) below]. The spin-orbit
interaction is implicitly included via the coefficients of the
$k$-dependent elements in the matrix. They contain matrix elements
of the spin-orbit interaction with respect to the cell-periodic wave
functions $u_{n\mathbf{k}}$ and can be evaluated theoretically.
\cite{Dresselhaus} However, a higher accuracy can often be obtained
from cyclotron resonance experiments, \cite{Hensel} and in practice
experimentally determined effective masses and Luttinger parameters
\cite{Luttinger} are inserted if they are available.

Cyclotron resonance experiments for a two-dimensional hole gas at a
modulation-doped GaAs/AlGaAs interface were performed by St\"{o}rmer
\emph{et al}. \cite{Stormer} The roughly triangular potential with a
strong Rashba effect gave rise to two clearly different masses
ascribed to the two components of the spin-split heavy-hole subband.
These results were in very good agreement with calculations with the
Luttinger-Kohn Hamiltonian, where the measurable energies of the
allowed Landau-level transitions were explicitly calculated.
\cite{Altarelli}

An alternative to the numerical solution of a multiband problem with
suitable boundary conditions is a process called "downfolding."
Starting from a multiband matrix, one can derive various $2 \times
2$ Hamiltonians. \cite{Lassnig,Silva,Pfeffer} The commonly used
Rashba term can be derived as the lowest-order term.
\cite{Winklerbook} For the electric field in the $z$-direction, it
becomes \cite{Bychkov}
\begin{equation}
H_R = \alpha (k_x \sigma_y - k_y \sigma_x)
\end{equation}
Here $\alpha$ is often used as an input parameter taken from
experiment. Several experiments (see, e.g., Refs.
\onlinecite{Richards}, \onlinecite{Nitta}, and
\onlinecite{Grundler}) have aimed at determining this Rashba
coefficient for different materials. Various theoretical expressions
have been derived for $\alpha$. A simple expression that we will use
as a reference is \cite{Silva}
\begin{equation}
\alpha = \frac{\hbar^2 e \Delta (2 E_g + \Delta)}{2 m E_g (E_g +
\Delta) ( 3 E_g + 2\Delta)} <\varepsilon>,
\end{equation}
where $<\varepsilon>$ is the expectation value of electric field in
the quantum well and the barriers.

The Rashba term is not really consistent with the principles of the
EFA. It is a kind of hybrid including both the wave vector $k$ and
the potential. The problem arises from the fact that the
\emph{s}-like conduction band gets its spin-orbit coupling from the
interaction with the valence bands, which are included in this
approximation among the "remote" bands. A further problem is that
Eq. (6) implicitly assumes that the electron subband has a
well-defined expectation value but as will be discussed below, the
two components of the spin-split subband (henceforth denoted spin
subbands) can have clearly different expectation values.

A special problem is how the interfaces of a quantum well should be
included. In an asymmetric quantum well the penetration of the wave
function into the left and right barriers becomes different. At a
first glance it seems natural to treat the complete conduction-band
profile as the relevant potential. Zawadzki and Pfeffer
\cite{Pfeffer} have included the conduction band offsets in
$<\varepsilon>$ [cf. Eq. (6)] and denoted it the "average electric
field." Using the fact that no force acts on a bound state it has
been argued \cite{Winklerbook,Pfeffer} that the contribution from
the interfaces would largely cancel that of the electric field in
the quantum well. However, in this respect it is important to
distinguish between the total wave function and the envelope
functions for which this argument does not necessarily hold. The
different effective masses in well and barrier, and spin dependent
boundary conditions make this "average electric field" yield a
non-zero but small contribution to the Rashba spin splitting.

Lassnig \cite{Lassnig} argued that the valence-band profile,
including the band offsets, determines the Rashba effect for
conduction electrons. This gives an interface contribution of the
same sign as that of the electric field in the quantum well and
barriers. This interface contribution has been evaluated
analytically by downfolding the other bands on the conduction band
and resulted in matrix elements of the step-like valence-band
edge.\cite{Pfeffer} However, this approach of treating the
interfaces like infinite electric fields seems dubious against the
discussion above, where it was pointed out that the interfaces are a
weak point in the EFA. A crucial nontrivial factor in the
downfolding procedure is how the order between differential
operators and spatially varying material parameters should be chosen
to be compatible with the Burt-Foreman boundary conditions.
\cite{Foreman93,Foreman97} So far downfolding approaches have
usually used the \emph{ad hoc} operator symmetrization that many
workers have abandoned in multiband approaches. We will analyze the
interface contribution below in our multiband approach using Eqs.
(5) and (6).

This interface contribution is fundamentally different from another
interface contribution where the microscopic structure at the
interface is taken into account as an additional source of inversion
asymmetry. The proper inclusion of this short-range potential in the
framework of envelope function theory is not trivial and has been
subject to debate.\cite{Foreman99} The effect is particularly strong
in $"$no-common-atom interfaces$"$ such as InAs/GaSb where the two
constituents have no atom in common.\cite{Krebs96} We have neglected
it in the present calculations.

Equations (5) and (6) predict a linear increase in the Rashba spin
splitting with the in-plane wave vector \emph{k}. Numerical results
\cite{Silva,Yang} give a Rashba splitting that is nonlinear and can
even be nonmonotonic. Yang and Chang \cite{Yang} have recently
published numerical and approximate analytical solutions to the
Rashba effect. By inserting the actual subband energy
$\varepsilon(k)$ instead of the bulk band-edge energies as in
simpler models, their analytical model qualitatively reproduces the
nonlinear $k$ dependence. However, the analytical model
overestimates the spin splitting. Their numerical results for an
In$_{0.53}$Ga$_{0.47}$As quantum well between
In$_{0.52}$Al$_{0.48}$As barriers agree very well with what we
obtain in our approach. The quantitative agreement for
Hg$_{0.74}$Cd$_{0.26}$Te/HgTe structures is less good but the
qualitative behavior of an increase followed by a decrease in the
spin splitting is reproduced. We find a similar nonmonotonic
behavior for an InGaSb/InAlSb structure with the same composition as
in this paper but with a uniform electric field. On the other hand,
for a GaAs/AlGaAs quantum well we find a monotonic but clearly
nonlinear behavior, in agreement with Ref. \onlinecite{Silva}. This
is consistent with the explanation \cite{Yang} that the nonmonotonic
$k$ dependence is caused by the increased energy separation to the
light-hole and split-off bands, an effect that is less pronounced in
GaAs with a fairly large band gap. It can be noted that the
dependence of $\Delta E$ on $k$ is similar to the dependence for
holes in Ref. \onlinecite{EPL} of $\Delta k$ at the optimal energy
on the electric field which was explained in a similar way.

We have started from the properties of the bulk materials and, via
the boundary conditions, obtained the subband dispersions. An
alternative approach is to solve the subband problem for $k = 0$ and
expand the wave functions for finite $k$ in eigenfunctions for $k =
0$ (the mini-$\mathbf{k} \cdot \mathbf{p}$ method). \cite{Broido}
This approach can be illuminating also for spin phenomena.
\cite{Winklerbook} For a modulation-doped interface it was found
that, however, the convergence with respect to number of basis
functions was surprisingly slow. \cite{Ando}

\section{Results}

\begin{figure}[!t]
\includegraphics{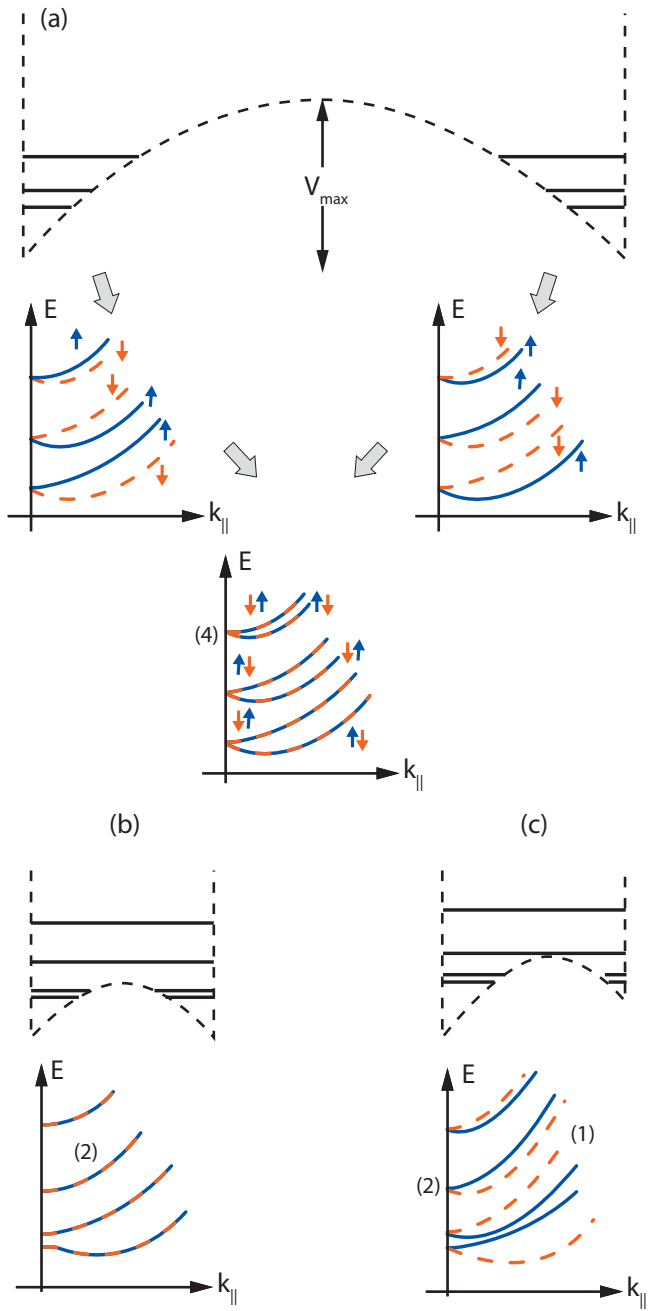}
\caption{(Color online) Schematic subband dispersions and
conduction-band profiles for \emph{n}-type MD quantum wells. The
numbers in parentheses denote the degeneracy of the subbands for $k
\ne 0$ and $k = 0$. In these figures the Dresselhaus splitting has
been ignored. (a) Here the well is wide enough that the two electron
gases at the interfaces can be considered as noninteracting. To the
left and the right we display the subband dispersions when the
interface regions are considered separately. The middle figure shows
the case when the quantum well is considered as a whole. (b)
Symmetric MD quantum well of intermediate width. The previous
fourfold degeneracy at $k=0$ partially lifted due to the interaction
between the electron gases. (c) Asymmetric MD quantum well. Due to
the overall asymmetry the degeneracy is lifted for $k \ne 0.$
\label{schematic}}
\end{figure}

We here consider an interesting system for studying spin effects, an
\emph{n}-type wide modulation-doped quantum well. Due to the
attraction to the ionized donors in the barriers, we obtain two
weakly interacting electron gases mainly localized to the interface
regions. The Rashba effect has often been studied experimentally and
theoretically in structures where the electric field is uniform or
close to uniform. In wide modulation-doped quantum wells the
electric field is strongly nonuniform [see Fig.~\ref{schematic}(a)].
In each interface region there is a strong electric field which,
according to Poisson's law, is proportional to the charge
transferred from the donors in the barrier to the quantum well. This
field is capable of producing a substantial Rashba effect.

The modulation-doped quantum well is a very versatile system. The
degree of interaction between the electron gases can be controlled
by the well width and the carrier concentration. The asymmetry can
be regulated by an applied bias or by choosing unequal spacer layer
widths.

In this paper we consider an 80 nm wide \emph{n}-type
In$_{0.74}$Ga$_{0.26}$Sb quantum well with In$_{0.7}$Al$_{0.3}$Sb
barriers. The growth direction is [001]. The input parameters are
given in Table I. Among the common III-V semiconductors InSb has the
strongest spin-orbit coupling. By mixing in Ga and Al, respectively,
we obtain a quantum well with an almost equally strong spin-orbit
coupling. This InGaSb/InAlSb system has recently been studied
experimentally by Akabori \emph{et al}. \cite{Akabori} Similar
effects should occur also for smaller well widths but the effects we
want to display become quite clear for a well width of 80 nm. We
here make a much more thorough analysis than in a recent preliminary
paper. \cite{APL}

We use the well established approach with an $8 \times 8$ matrix
approach with a minor modification (see Table Caption I) of the
matrix given as Table C.8 in Ref. \onlinecite{Winklerbook}. We apply
the Burt-Foreman boundary conditions,
\cite{Burt,Foreman93,Foreman97} and to avoid spurious solutions, we
use a quadrature method in which unphysically large \emph{k} values
do not enter.\cite{Rossler} The subband problem is solved
self-consistently in the Hartree approximation.

For reference we first consider a symmetric quantum well. If we, for
the moment, ignore the Dresselhaus effect, we have a ground state
with a symmetric wave function and a small energy separation to an
excited state with an antisymmetric wave function. If the
modulation-doped quantum well is wide enough that the electron gases
can be considered as noninteracting, we seem to have a paradox. Each
electron gas is in a strongly asymmetric potential and a strong
Rashba effect can be expected. On the other hand, if the quantum
well is considered as a whole, the potential is symmetric and a
twofold spin degeneracy should result [see Fig.~\ref{schematic}(a)].
This was sorted out previously \cite{PRB} for a \emph{p}-type
quantum well, where the Rashba effect can be made several orders of
magnitude larger than for electrons (cf. Ref. \onlinecite{EPL}).
However, the qualitative features are the same for the \emph{n}-type
quantum well. We recapitulate here the essentials and refer to Ref.
\onlinecite{PRB} for details.

First it should be noted that the signs of the electric fields at
the interfaces are opposite to each other. If we label the upper
spin subband at the left interface by spin up, the corresponding
spin subband at the right interface should be labeled spin down.
Looking now at the whole quantum well with two electron gases, the
lower spin subband has equal amounts of spin up and spin down, and
the twofold spin degeneracy expected for a symmetric potential
prevails [lower subband structure in Fig.~\ref{schematic}(a)]. This
implies that what looks like a spin-split subband in the single
interface case should actually be considered here as two separate
subbands. At $k = 0$ we have a four-fold degeneracy because we have
two spin directions and two electron gases. If we now consider a
narrower but still symmetric quantum well such that the electron
gases start to interact, the main and somewhat unexpected effect is
that the degeneracy at $k = 0$ is partially lifted, and for all
$k$-values we have two closely spaced subbands, each with a twofold
spin degeneracy [Fig.~\ref{schematic}(b)].

\begin{table}[!t]
\caption{Input parameters used in the present work for an
In$_{0.74}$Ga$_{0.26}$Sb quantum well surrounded by
In$_{0.7}$Al$_{0.3}$Sb barriers. We give the $"$true$"$ Luttinger
parameters although modified Luttinger parameters are included in
the $8 \times 8$ matrix.\cite{Winklerbook} The conduction-band
offset $\Delta E_c$ and valence-band offset $\Delta E_v$ between the
two materials are also given. Mass parameters are given in units of
the free-electron mass. Energies are given in electron volts. The
coefficients $C$ and $B^+_{8v}$ describing the Dresselhaus effect
are given in eV {\AA} and eV {{\AA}$^2$}, respectively. We use the
reasonable approximations $B_{7v} \approx B^+_{8v}$ and $B^-_{8v}
\approx 0$ that are justified by Table 6.3 in
Ref.\onlinecite{Winklerbook}. Finally we give the static dielectric
constant $\epsilon$.}
\begin{tabular}{llll} \\ \hline \hline
Parameter & InGaSb & & InAlSb \\ \hline
$m_e$ & 0.01595 &  & 0.04221 \\
$\gamma_1$ & 23.87 &  & 11.652 \\
$\gamma_2$ & 10.01 &  & 4.110 \\
$\gamma_3$ & 11.34 &  & 5.388 \\
$E_p$ & 24.27 &  & 21.92 \\
$E_g$ & 0.3068 &  & 0.790 \\
$\Delta$ & 0.777 &  & 0.717 \\
$C$ & -0.006715 &  & -0.006524 \\
$B^+_{8v}$ & 20.596 &  & 7.21 \\
$\Delta E_c$ &  &  0.3344 &  \\
$\Delta E_v$ &  &  0.1487 &  \\
$\epsilon$ & 17.17 &  & 16.00 \\
 \hline \hline
\end{tabular}
\end{table}

This effect can possibly be related to the splitting recently
obtained by Bernardes \emph{et al.} \cite{Bernardes} who derived
some off-diagonal matrix elements between wave functions of even and
odd parity. This gives rise to a splitting qualitatively similar to
the splitting described above when the electron (or hole) gases
start to interact. However, the mechanism for the splitting derived
in Ref. \onlinecite{Bernardes} is the spin-orbit interaction. Since
well widths and other input parameters are not given in this paper,
we cannot investigate if this effect is implicitly included in our
multiband approach.

\begin{figure}[!t]
\includegraphics{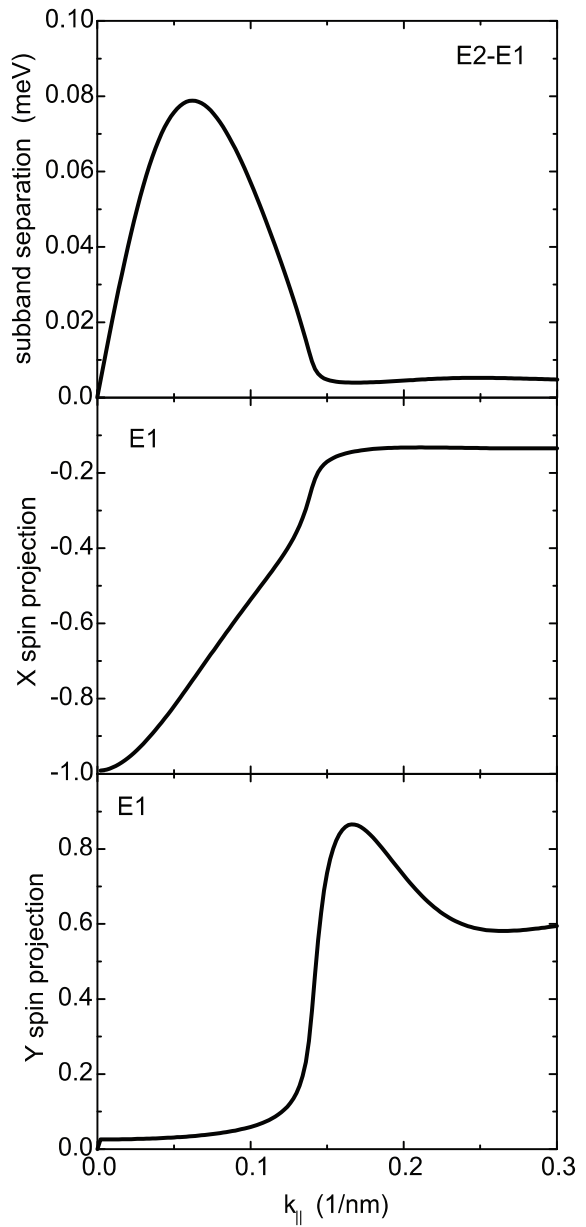}
\caption{(a) Energy subband splitting for an 80 nm \emph{n}-type
modulation-doped symmetric InGaSb quantum well. The wave vector is
in the [10] direction in the two-dimensional Brillouin zone. [(b)
and (c)] $k$ dependence of the $x$ and $y$ components, respectively,
of the spin vector. Here both the spacer layer widths are 45 nm and
the electron density is $6.8 \times 10^{11}$ cm$^{-2}.$\label{symm}}
\end{figure}

The two-fold spin degeneracy becomes lifted for finite $k$ if the
potential of the quantum well is made asymmetric
[Fig.~\ref{schematic}(c)]. This is the case treated in this paper.
The behavior described above is confirmed by numerical calculations.
\cite{PRB}

\begin{table*}[!t]
\caption{Spin splittings for different electric fields in various
approximations. $\varepsilon_{ave}$ is the voltage across the
quantum well divided by the well width and $\Delta E_{ave}$ is the
energy spin splitting obtained by inserting this electric field
into Eq. (6). $\langle \Delta E \rangle_i^{excl}$ are the results
when the expectation value of the electrostatic field in the
layers excluding interface contributions (see text) averaged over
filled states for spin subband $i$ is inserted into Eq. (6).
$\langle \Delta E \rangle_i$ are the corresponding results with
inclusion of interface contributions from the valence band offsets
(Eq.(7)). $\Delta E_{num}$ is our numerical result with the spin
splittings evaluated at the Fermi wave vector $k_F \approx 0.13$
nm$^{-1}$. The last row corresponds to the situation in Fig. 3.}
\begin{tabular}{lllllll} \\ \hline \hline
$\varepsilon_{ave}$ & $\Delta E_{ave}$ & $\langle \Delta E
\rangle_{1\downarrow}^{excl}$&$\langle \Delta E
\rangle_{1\uparrow}^{excl}$&$\langle \Delta E
\rangle_{1\downarrow}$&$\langle \Delta E \rangle_{1\uparrow} $ & $\Delta E_{num}$ \\
(kV/cm) & (meV) & (meV) & (meV) & (meV) & (meV) & (meV) \\ \hline
0.297 & 0.012 & 0.862 & -0.292 & 1.384 & -0.495 & 0.720 \\
1.242 & 0.052 & 1.112 & 0.968 & 1.768 & 1.491 & 2.158 \\
3.012 & 0.135 & 1.357 & 1.328 & 2.145 & 2.040 & 2.498 \\
4.137 & 0.190 & 1.485 & 1.459 & 2.336 & 2.231 & 2.646 \\
\hline \hline
\end{tabular}
\end{table*}

We next include the Dresselhaus terms in the matrix but keep the
quantum well potential symmetric. The Dresselhaus effect becomes
rather different in a wide modulation-doped quantum well compared to
a square well. \cite{Winklerbook} In Fig.~\ref{symm}(a) we display
the energy spin splitting as a function of wave vector. It is seen
that it first rapidly increases but then decreases and for larger
wave vectors (approximately above the Fermi wave vector $k_F$), it
stays rather constant at a low value. Our results imply that the
change in spin splitting between a symmetric and an asymmetric wide
modulation-doped quantum well normally is dominated by the Rashba
effect. In Figs.~\ref{symm}(b) and \ref{symm}(c) we show the $k$
dependence of the $x$ and $y$ components of the expectation value of
the spin vector.\cite{Winklerbook} The absolute value of the $x$
component decreases rapidly and becomes small above $k_F$. This is
in contrast to a square well where it stays constant.
\cite{Winklerspin} The $y$ component has the reverse behavior: small
for small $k$ values and increases rapidly near $k_F$. Thus the spin
direction changes from the $x$ direction to the $y$ direction as $k$
increases along the [10] direction. For sufficiently large \emph{k}
we find a localization of the wave functions to one of the interface
regions, similarly to what is shown below for the Rashba effect.

In a quantum well with small asymmetry, the Rashba splitting can be
comparable to the splitting between the symmetric and the
antisymmetric states. In addition to a gradual transfer of
wave-function amplitude to one of the interface regions, there will
be interesting anticrossing phenomena, especially when $k$ is in the
[11] direction. \cite{Basel,Genova} In Table II we show the spin
splittings for some values of the bias in various cases. Especially
for small bias the expectation values can be quite different and
even have opposite signs for the two spin subbands. This presents a
fundamental problem in applying Eqs. (5) and (6) for more complex
situations. Many downfolding approaches lead to expectation values
of the electric field with respect to a subband rather than a spin
subband. If one wants to treat this effect perturbatively, it would
be appropriate to apply degenerate perturbation theory. However,
since we find that a small perturbation can give rise to a
substantial effect on the wave functions, the use of perturbation
theory appears dubious in the present case. The anticrossing
phenomena including the contrasting results for $k$ in the [11]
direction have been examined more closely
elsewhere.\cite{Basel,Genova}

\begin{figure}[!t]
\includegraphics{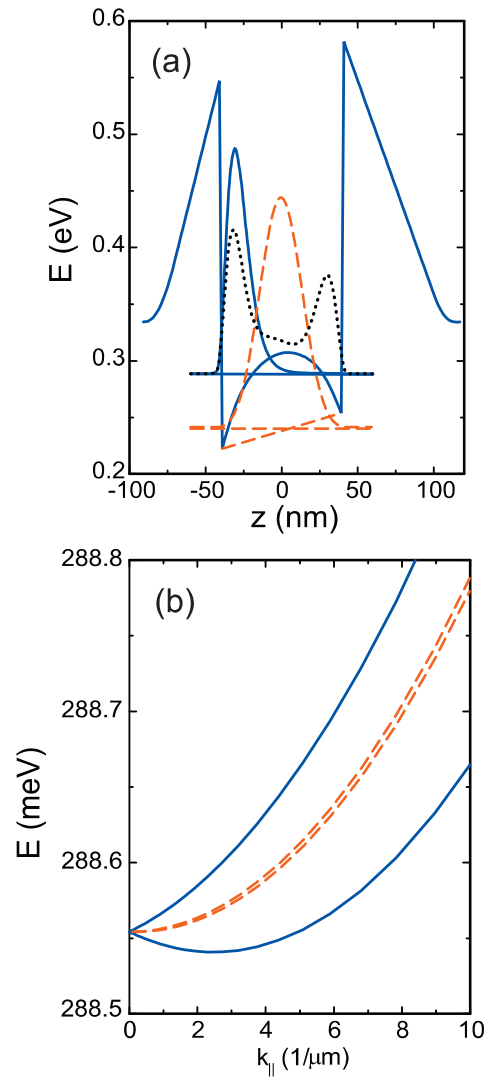}
\caption{(Color online) (a) Potential, squared wave function and
charge density, and (b) subband dispersion along the [10] direction
for the lowest subband pair in an 80 nm InGaSb quantum well. The
quantum well bias (potential difference between the interfaces) is
33 mV. Dashed lines: uniform electric field; solid lines:
modulation-doped quantum well with an electron density of $6.8
\times 10^{11}$ cm$^{-2}$. For the latter case the dotted line shows
the charge distribution. \label{enhanced}}
\end{figure}

In this paper we focus on the case with a fairly small but
sufficiently large asymmetry that each wave function becomes almost
completely localized to one of the interface regions. In the present
case a bias over the quantum well [henceforth denoted quantum well
bias (QWB)] of 33 mV is sufficient to reach this situation. It
corresponds to a rather small average electric field of 4.1 kV/cm
(last row in Table II). It gives an energy separation of 9.2 meV at
$k = 0$ between the lowest and next lowest subband pairs. It is
shown in Fig.~\ref{enhanced}(b) that it gives a spin splitting that
is an order of magnitude larger than for the same uniform electric
field. By comparing columns 2 and 7 in Table II, we can observe for
different biases how the spin splitting is enhanced in
modulation-doped quantum wells compared to undoped quantum wells
with the same QWB. For the largest (last row) and the smallest
biases (first row), we have an enhancement by a factor of 14 and 60,
respectively.

We thus have a modified and very efficient mechanism to apply a
moderate QWB, and take advantage of the much stronger built-in
electric field to obtain a substantial Rashba splitting. A
qualitative explanation of the enhancement can be seen in
Fig.~\ref{enhanced}(a) where the ground-state wave function has
become localized to one of the interface regions. There the electric
field becomes quite strong and it is this local field that
determines the size of the spin splitting.

We note in Table II than we can reach a spin splitting of 2.6 meV at
a wave vector of 0.13 nm$^{-1}$. According to Eq. (5) this
corresponds to value as large as 200 meV {\AA} for the Rashba
parameter $\alpha$. This compares well in comparison to
experimentally determined $\alpha$ values.
\cite{Richards,Nitta,Grundler} In addition, it is essential that in
our case this $\alpha$ value is reached with a moderate bias.

We will now investigate if we can reproduce this strong enhancement using the
common Rashba model together with Eq. (6). As a first step we insert into
Eq. (6) the expectation value of the electric field in the well and barriers
ignoring any interface contributions (columns 3 and 4). It can be expected to
be enhanced by the localization of the wave function. In the last row of Table II
the spin subbands have almost the same expectation value. However, this procedure
gives a clearly smaller Rashba splitting than obtained in our numerical
calculations. We have found that inclusion of BIA has a small effect
on the results in Table II.

As a next step it is natural to examine if the discrepancy can be
explained by interface contributions. After solving the multiband
problem in our approach, it is straightforward to evaluate the
expectation value of the electric field in the layers and that of
the steps at the interfaces separately. If the Rashba model Eqs. (5)
and (6) is to be used, the most reasonable approach seems to be to
follow Ref. \onlinecite{Lassnig} and take the expectation value of
the valence-band profile. The interface contributions involve the
derivative of the discontinuities in the valence bands and yield
Dirac delta functions in the integrals that nevertheless can be
evaluated. The result becomes
\begin{equation}
<\varepsilon>_{interfaces} = \Delta E_v (|\psi(-a)|^2
-|\psi(a)|^2)/e,
\end{equation}
where the interfaces are taken at $z = \pm a$. It is seen in Table
II that the expectation values are increased by about 50$\%$ when
the contribution from the interface steps is added. This is
compatible with the analytical results by Yang and Chang \cite{Yang}
who concluded that the interface contribution to the Rashba
splitting was typically a factor of two smaller than that of the
electric field in the quantum well.

We have also calculated the expectation value of the conduction-band
profile to obtain the "average electric field".\cite{Pfeffer} The
contribution from the conduction-band offsets is of opposite sign
compared to that from the electric field in the quantum well and
results in a net contribution of 10$\%$-15$\%$ of our numerical
value. In Ref. \onlinecite{Pfeffer} a contribution of only 3$\%$ was
obtained.

To examine the interface effect we have also replaced the barrier
material by In$_{1-x}$Al$_x$Sb of different compositions. This
changes the conduction- and valence-band offsets, the effective
electron mass, and the nonparabolicity. If the interface
contribution were very sensitive to these parameters, this should
give a clear effect. However, in our calculations we only find a
minor change. For example, if we change the barrier material from
In$_{0.7}$Al$_{0.3}$Sb to In$_{0.5}$Al$_{0.5}$Sb (ignoring
introduced strain), we obtain a change of the spin splitting by
about 1$\%$.

The failure to reproduce our numerical results using Eqs. (5) and
(6) and various expectation values raises the question if more
elaborate downfolding approaches can yield better agreement or if
approaches using expectation values simply are insufficient. A more
comprehensive comparison between various multiband and downfolding
approaches will be published elsewhere. It is conceivable that two
potentials having the same expectation value of the electric field
can yield different Rashba spin splittings and that the spatial
variation in the electric field must be taken into account in a
multiband approach rather than basing the calculations on some kind
of expectation values.

\section{Implications for spintronic devices}

The strong enhancement of the Rashba splitting described in
Fig.~\ref{enhanced} due to modulation doping can be expected to have
important implications for several spintronic devices based on the
Rashba effect. For the moment we focus on one of the best known
spintronic devices, the spin transistor proposed by Datta and
Das,\cite{Datta} including proposed modifications of it. We will
return to the problems encountered to make it function and first
address the question: If it can be made to function, does it have
the potential to become competitive with state-of-the-art
conventional transistors? Then it is not only essential that one can
achieve a large wave vector splitting $\Delta k$ of a spin-split
subband but also that it can be done with a small bias. As a
benchmark for the performance, we choose the switch energy for Si
MOSFETs where 3 aJ has been projected. \cite{ITRS}

We have previously \cite{Gvozdic} approximated the switch energy for
\emph{n}-type and \emph{p}-type spin transistors by $C V^2$, where
$C$ is the capacitance of a QW structure surrounded by two gates and
$V$ is the applied bias between them. (We have included here turning
on and off of the device, which cancels a factor $1/2$). We then
concluded that \emph{n}-type spin transistors with the original
design would have problems in becoming competitive with conventional
transistors unless fundamentally new ideas were presented.

A similar conclusion was independently drawn by Bandyopadhyay and
Cahay. \cite{Cahay} They assumed that a spin transistor must be
based on a one-dimensional channel and that only the lowest
one-dimensional subband should be filled. However, this resulted in
an anomalously small carrier density, $3 \times 10^{10}$ cm$^{-2}$
or $3 \times 10^5$ cm$^{-1}$. This made their comparison very
unfavorable for the spin transistor.

A more recent comparison with conventional transistors has been made
by Hall and Flatt\'{e}\cite{Hall} for a modified spin transistor. It
is not based on the Rashba effect but rather on gate-induced spin
relaxation. A crucial factor in their approach seems to be
efficiency of the gate-induced spin relaxation compared to other
spin-relaxation mechanisms. Their comparison was quite favorable for
the spin transistor. They estimated a switch energy of 0.5 aJ which
is similar to what we find below for our modified spin transistor.
The performance of this transistor has been subject to some
controversy concerning the need for very efficient spin
injection.\cite{Debate} Since the Hall-Flatt\'{e} spin transistor is
based on another mechanism than the spin precession considered here,
it is beyond the scope of the present paper to enter this debate.

Utilizing the built-in electric field in the modulation-doped
quantum well, one can achieve a given $\Delta k$ with a QWB that is
an order of magnitude smaller than with a uniform electric field. If
we only consider the lowest spin subband pair and follow the
approach of Ref. \onlinecite{Gvozdic}, we obtain a switch energy of
0.4 aJ in the modulation-doped case and 35 aJ in a spin transistor
with the same length and uniform electric field. The former figure
compares very well with present state-of-the-art transistors. Thus
the utilization of the mechanism proposed by us could make a
substantial difference for the competitiveness of spin transistors.

We have calculated the additional contribution to the switch energy
from the redistribution of carriers in the QW, taking the
$k_{\|}$-dependent wave functions into account but found that it
only increases by about 20 $\%$.

A complication with our design is that the second subband pair with
the opposite sign of $\Delta k$ and spin precession direction is
also filled. This does not prevent the possibility that the spins at
the two interfaces can have made a precession by the angle $\pi$ but
in opposite directions on the arrival to the drain where the
transmission becomes low.

It has been demonstrated that one can contact the electron gases in
a double quantum well structure separately. \cite{Eisenstein} It
seems feasible that also the interface regions of a wide
modulation-doped quantum well can be contacted separately which
opens up interesting possibilities occurring from the controllable
properties of modulation-doped quantum wells.

One can envision practical problems to create a perfectly symmetric
quantum well structure corresponding to the on state of a spin
transistor. One possibility is a double-gate structure in which the
total carrier concentration and the asymmetry can be controlled
separately. In Ref. \onlinecite{Papadakis} the back gate voltage was
of the order 100 V, which is not very practical for devices. An
alternative design \cite{Gvozdic} is to have a heavily doped
semiconductor layer just below the quantum well structure. In this
way a larger fraction of the applied voltage falls over the quantum
well.

We now turn to the problem of making a spin transistor function,
possibly with some modification of the original idea. \cite{Datta} A
fundamental problem is that the Rashba effect can be described in
terms of an effective magnetic field that is perpendicular to both
the electric field and the direction of motion for the
carrier.\cite{Winklerbook} Even spin-independent scattering leads to
a change of the direction of the velocity and thus the axis of the
spin precession. It can also be difficult to inject all the carriers
in the same direction. In the case we consider in
Fig.~\ref{enhanced}, we obtain a precession length $L = \pi/\Delta k
\sim 1 \mu$m. Ballistic transport over such a distance requires
rather low temperatures. An idea with the purpose of balancing the
Rashba and Dresselhaus effect \cite{Schliemann} by setting $\alpha =
\beta$ in the linearized model makes diffusive transport possible
but at the price of a substantial transmission in the off state.
One-dimensional channels have been proposed in which the carriers
are more or less forced to move in the same direction. As mentioned
above the small energy separation between the one-dimensional
subbands leads to multi-mode transport for realistic carrier
densities. This is not a prohibitive problem as has been
demonstrated by {\L}usakowski \emph{et al}.\cite{Lusakowski}

A fundamental problem is that in the approach with the Rashba term,
which is a reasonable approximation for electrons in an undoped
quantum well, the Rashba splitting $\Delta k$ is proportional to the
Rashba coefficient $\alpha$ but the spin decoherence rate becomes
proportional to $\alpha^2$ (Ref. \onlinecite{Zutic}). A large
$\alpha$ is beneficial for a rapid spin precession and corresponds
to a short gate length in a spin transistor but this advantage is
thus offset by the shorter spin decoherence time.

An alternate approach has been presented by Bandyopadhyay and Cahay.
\cite{Bandy} Instead of relying on the Rashba effect, they propose
using the Dresselhaus effect in a structure with a split gate and a
parabolic potential. The main reason was to avoid an in-plane
magnetic field in the semiconductor from the magnetized source and
drain. However, the main requirement for spin precession is that
their magnetization is perpendicular to the effective magnetic field
in the channel. Thus it can be either along the channel (as drawn in
Ref. \onlinecite{Bandy}) or perpendicular to the layers (as drawn in
Ref. \onlinecite{Winklerspin}). The mechanism in the transistor
based on the Dresselhaus effect is changing the bias of the split
gate and then it is assumed that the curvature of the parabolic
potential changes. Numerical calculations \cite{Laux} have indicated
that, however, the effect of changing the bias is mainly that the
potential becomes flatter in the middle when the channel starts to
fill while the curvature of the side walls does not change much.

An alternative that has not been given much attention so far is a
\emph{p}-type spin transistor. With a suitable design we have shown
that one can obtain a large $\Delta k$ with an electric field as
small as 2 kV/cm (Ref. \onlinecite{EPL}). The corresponding
precession length is only 40 nm and the possibility of having
ballistic transport over such a short distance clearly seems
feasible. The strong spin-orbit interaction including its dependence
on a gate voltage has recently been demonstrated for a \emph{p}-type
GaAs/AlGaAs heterostructure by Grbi\'{c} \emph{et al.}\cite{Grbic}
According to our calculations \cite{EPL} even stronger spin
splitting can be achieved for higher hole densities. Furthermore,
the strong anisotropy of hole subbands can possibly be utilized to
get a preferred direction of motion without lateral confinement. For
holes there is no simple relation between spin precession and spin
decoherence rates. Estimates based on experimental determinations
indicate that the spin decoherence time can be much longer than the
transit time. Because of the strongly nonparabolic hole subbands and
their mixed heavy-hole and light-hole character, rather cumbersome
numerical calculations appear necessary for a more accurate
prediction of the transport properties. For small $k$ analytical
expressions proportional to $k^3$ for the Rashba splitting in
heavy-hole subbands have been derived. \cite{Winklerbook} However,
it has been found that the largest spin splittings occur beyond the
range of validity of this model. \cite{EPL}

A relevant question is if one can combine the superefficient Rashba
effect for holes with the enhancement in modulation-doped quantum
wells presented here. However, we have shown that for \emph{p}-type
spin transistors the largest Rashba splitting is obtained for quite
small electric fields ($\sim$ 5 - 10 kV/cm) while the effect of
modulation doping is to apply a small bias to utilize the built-in
electric field of the order $50 - 100$ kV/cm for which the Rashba
effect for holes is reduced.

A well-known problem is that the conductivity mismatch between metal
and semiconductor severely limits the spin injection
efficiency.\cite{Schmidt} A proposal by Rashba is having tunnel
barriers between the metal contacts and the
semiconductor.\cite{Rashba} A fundamental problem recently pointed
out by Fert \emph{et al}. \cite{Fert} is that this decreases the
transmission coefficient and increases the dwell time such that it
can become long compared to the spin dephasing time in
semiconductor-based spin transistors. They instead proposed using
carbon nanotubes. It is beyond the scope of the present paper to
evaluate the competitiveness of semiconductors vs. carbon nanotubes
for spintronic devices. However, we would like to point out that
this problem occurs for injection from a spin-polarized contact but
other solutions in the form of spin filters have been proposed. A
particularly interesting idea is to put a magnetic layer on top of a
layered semiconductor structure such that the in-plane fringe fields
act as a spin filter.\cite{Xu} The appealing aspect of this solution
is that current flows in the channel below the metal without passing
any interfaces where the spin polarization can be reduced.

\section{Discussion and conclusions}

We have implicitly assumed coherence of the wave function across the
80 nm QW with a high and broad barrier in the middle. Whether this
coherence actually prevails should depend on the sample quality.
This system with our predicted effects seems ideal for further
studies of this fundamental problem.

The self-consistent calculations have so far been performed in the
Hartree approximation. For studies of spin properties it is
conceivable that exchange and correlation can give significant
effects, especially in anticrossing situations. This is planned to
be examined in future publications.

In our multiband approach the well established Burt-Foreman boundary
conditions \cite{Burt,Foreman93,Foreman97} are behind any interface
contribution. The exact relation behind this approach and what is
obtained by folding down the adjacent bands to the conduction band
as in Refs. \onlinecite{Pfeffer} and \onlinecite{Lassnig}, is not
trivial and remains to be analyzed. Our approach is in our opinion
more sound than one-band approaches that are based on approximations
whose accuracy is difficult to determine. Furthermore, the
analytical expressions \cite{Lassnig,Pfeffer} are based on
consideration of the interfaces, where the EFA has its main weakness
and where the actual gradual but rapid potential variation near an
interface is replaced by a sharp step. Especially when the
interfaces give substantial contributions, it is likely that
different operator orderings can influence the results considerably.
Calculations with downfolding procedures for wide modulation-doped
quantum wells and comparison with the present results would be
valuable in evaluating how close the results of these approaches are
in a non-trivial case such as this.

It has recently been predicted that the spin Hall effect can be
strongly enhanced at a subband anticrossing in a bilayer
system.\cite{Jin} There the potential was not specified but the
Rashba coefficients were allowed to differ in the two layers. For
further investigations of this effect modulation-doped quantum wells
seem useful due to the possibility of controlling the degree of
interaction between the two electron gases and each of the interface
fields.

In conclusion we have analyzed the foundations of the envelope
function approximation and concluded that, while the Dresselhaus
effect should be included as $k$-dependent terms in a matrix, the
proper inclusion of the Rashba effect is adding the macroscopic
potential along the diagonal in a multiband approach. This has given
good agreement with experiment for two-dimensional hole gases.
\cite{Bangert,Stormer,Altarelli} The commonly used Rashba term (5)
is a hybrid including both potential and $k$. The proper derivation
of such a term within the framework of the EFA with proper boundary
conditions \cite{Burt,Foreman93,Foreman97} deserves to be examined
more closely.

For symmetric wells with Dresselhaus effect only, we find
interesting effects in a modulation-doped quantum well that are
qualitatively different from those in a square well.

We have found that, with a non-uniform electric field, insertion of
some kind of expectation value or other average into Eq. (6)
underestimates the Rashba splitting. Furthermore, this expectation
value is not always well-defined for a subband because it can differ
substantially between its spin-split components. We have found that
the contribution from the interfaces is about half of that from the
electric field in the layers for the potential we have considered.

We have demonstrated a very efficient switching mechanism of the
Rashba splitting in wide modulation-doped quantum wells. One can use
a bias corresponding to a moderate average electric field and still
get a Rashba splitting typically enhanced by an order of magnitude
due to the built-in local electric field in the interface region.
The switching mechanism is based on localization of each wave
function to one interface region with a barely sufficient bias. A
switching mechanism based on anticrossing in slightly asymmetric
quantum wells \cite{Basel,Genova} is not included here but will be
examined further elsewhere.

The enhancement of the efficiency of the Rashba effect should be
valuable for different spintronic devices. We here have focused on
spin transistors of the type proposed by Datta and Das. \cite{Datta}
With our modification we find that it can get a potential to
outperform conventional transistors. We have also discussed some
remaining obstacles to make such spin transistors function. \\

D.M.G. wishes to acknowledge the hospitality of the School of
Information and Communication Technology of the Royal Institute of
Technology where some of this work was carried out. Funding has been
received from the Serbian Ministry of Science.

\end{document}